\documentclass[runningheads]{llncs}
\usepackage[T1]{fontenc}
\usepackage{graphicx,verbatim}
\usepackage{multirow}
\usepackage{diagbox}
\begin{document}
\title{Radiomics and Clinical Features in Predictive Modeling of Brain Metastases Recurrence}
%
\author{
Inês Faria\and
Matheus Silva\and
Crystian Saraiva\and
José Soares\and
Victor Alves
}

\authorrunning{Faria et al.}

\institute{
University of Minho, Braga, Portugal\\
\email{a95494@alunos.uminho.pt}
}

\maketitle              
\begin{abstract}
Brain metastases affect approximately 20–40\% of cancer patients and are commonly treated with radiotherapy or radiosurgery. Early prediction of recurrence following treatment could enable timely clinical intervention and improve patient outcomes. This study proposes an artificial intelligence–based approach for predicting brain metastasis recurrence using multimodal imaging and clinical data.

A retrospective cohort of 97 patients was collected, including Computed Tomography (CT) and Magnetic Resonance Imaging (MRI) acquired before treatment and at first follow-up, together with relevant clinical variables. Image preprocessing included CT windowing and artifact reduction, MRI enhancement, and multimodal CT–MRI registration. After applying inclusion criteria, 53 patients were retained for analysis. Radiomics features were extracted from the imaging data, and delta-radiomics was employed to characterize temporal changes between pre-treatment and follow-up scans. Multiple machine learning classifiers were trained and evaluated, including an analysis of discrepancies between treatment planning target volumes and delivered isodose volumes.

Despite limitations related to sample size and class imbalance, the results demonstrate the feasibility of radiomics-based models, namely ensemble models, for recurrence prediction and suggest a potential association between radiation dose discrepancies and recurrence risk. This work supports further investigation of AI-driven tools to assist clinical decision-making in brain metastasis management.


\end{abstract}
\section{Introduction}

Cancer remains a major cause of mortality worldwide, responsible for millions of deaths annually. This condition is marked by the uncontrolled proliferation of abnormal cells, which can form tumors and metastasize to distant organs. Notwithstanding the noteworthy advancements in the domains of diagnosis and treatment, the global burden of cancer persists at a considerable level \cite{ref1}. Recent technological advancements, particularly in the domain of radiation therapy, have facilitated the precise delivery of therapeutic beams to tumors while minimizing exposure to surrounding healthy tissue. Concurrently, the integration of Artificial Intelligence (AI), Machine Learning (ML), and Deep Learning (DL) has further accelerated advances in the field of neuro-oncology research.

Artificial Intelligence (AI) refers to computational approaches that replicate or support tasks traditionally requiring human intelligence, such as decision-making, problem-solving, and natural language processing. Machine learning (ML) is a pivotal component of artificial intelligence (AI) due to its capacity to enhance the efficacy of algorithms through iterative learning from data, leading to a reduction in errors, and the ability to adapt to novel patterns.A subset of machine learning (ML) known as deep learning (DL) utilizes multi-layered neural networks that are capable of modeling highly complex, non-linear relationships and recognizing subtle patterns within large, heterogeneous datasets \cite{ref2,ref3}.

In recent years, the field of neuro-oncology has been increasingly influenced by technological advancements. Machine learning (ML) has demonstrated significant potential in supporting tumor management across its continuum. This continuum encompasses early diagnosis and risk stratification, prognosis, and individualized treatment planning. The primary mechanism through which ML supports tumor management is by leveraging advanced medical imaging \cite{ref4}. AI applications have been shown to complement and extend clinicians' expertise in several ways. These include the optimization of workflows, the provision of precise volumetric delineation of tumor size, the enablement of efficient analysis of large-scale imaging repositories, and the uncovering of clinically relevant patterns that may not be readily discernible to the human eye \cite{ref5}. This enhancement encompasses a range of domains, including the optimization of workflows, the provision of precise measurements such as volumetric delineation of tumor size, the analysis of substantial volumes of medical imaging data, and the identification of patterns that may not be readily apparent \cite{ref4}. 
The integration of artificial intelligence, particularly deep learning and radiomics, has significantly augmented the efficacy of these optimizations on this particular disease \cite{ref6}. This integration is being implemented in numerous domains, including drug discovery, medical diagnostics and imaging, remote patient care, risk management, hospital assistants, and virtual assistants.

Brain Metastases (BMs) represent the most prevalent form of cancer within the central nervous system (CNS), often inducing neurological complications in numerous cancer patients. The prevalence of brain metastases ranges from 20\% to 40\% among cancer patients, with a higher incidence observed in certain cases \cite{ref7}.

The incidence of brain metastases varies considerably depending on the histology and biological characteristics of the primary tumor. Among the most prevalent cancers that metastasize to the brain are those of the lung, breast, and skin (particularly melanoma). The discrepancy in reported incidence is influenced by a multitude of factors, including but not limited to tumor biology and demographic trends. The aging of the global population, for instance, has been identified as a contributing factor to the overall increase in cancer cases \cite{ref8}. Metastatic disease is characterized by the dissemination of malignant cells from the primary tumor to distant organs. In the context of brain metastases, circulating tumor cells invade the central nervous system, most frequently localizing within the cerebral hemispheres or, less commonly, the cerebellum, where they proliferate and form secondary tumor masses \cite{ref9}.

Magnetic Resonance Imaging (MRI) is the primary diagnostic modality for brain metastases, as it provides detailed information on tumor location, size, biological characteristics, and associated mass effect \cite{ref10}. The utilization of Computed Tomography (CT) in subsequent phases of treatment planning is a common practice. The management of newly diagnosed brain metastases is complex and may involve multimodal approaches, including surgery, radiotherapy, and stereotactic radiosurgery \cite{ref7}.

Radiotherapy, in particular, is a highly personalized treatment modality, tailored to the unique clinical and anatomical characteristics of each patient. The delivery of this intervention follows a structured process designed to maximize therapeutic effectiveness and minimize risks. The process is generally comprised of five stages: initial patient assessment, simulation, treatment planning, treatment delivery with rigorous quality assurance, and post-treatment follow-up \cite{ref19}.

The management of brain metastases can involve several therapeutic approaches, most notably Stereotactic Radiotherapy (SRT), Whole-Brain Radiotherapy (WBRT), and Stereotactic Radiosurgery (SRS) \cite{ref7}.

Despite the enhanced precision and expanded clinical applicability of these techniques, treatment response exhibits variability among metastatic lesions. The assessment of therapeutic effectiveness frequently necessitates protracted follow-up intervals. Imaging plays a pivotal role in this evaluation, offering insights into treatment response and enabling the prediction of clinical outcomes. This information is imperative for the guidance of therapeutic decision-making and the refinement of subsequent treatment planning.

A growing body of research has investigated the use of Artificial Intelligence (AI) to predict metastatic progression following treatment, drawing on a wide range of features, including clinical data, radiomics, semantic descriptors, and deep learning–derived representations. These methodologies have proven valuable in predicting treatment outcomes, thereby contributing to improved therapeutic effectiveness. As a result, Artificial Intelligence is reshaping medical practice by making treatments more efficient, accurate, precise, and of higher quality, being able to identify  complex patterns in data and conclude clinical diagnoses. The objective of this study is to investigate the feasibility of integrating radiomics features derived from treatment imaging with clinical attributes to predict brain metastases recurrence.

\section{Materials and Methods}

The present chapter focuses on the development and evaluation of predictive models for recurrence
in patients with brain metastases. A comprehensive processing pipeline was implemented, integrating clinical, radiological, and treatment-related variables. This pipeline encompassed several key steps, including patient data anonymization, image registration, and feature extraction. Machine learning techniques were subsequently applied to estimate patient-specific recurrence risk. The primary objective of this study is to compare the performance of the developed models, identify the most effective one, optimize its parameters, determine the most relevant predictive features, and evaluate their potential to support clinical decision-making. The chapter is structured to present the materials, methods, and results within the frame work of personalized patient management. The following figure illustrates the overall workflow developed
throughout this study.

\begin{figure}
    \centering
    \includegraphics[width=0.60\linewidth]{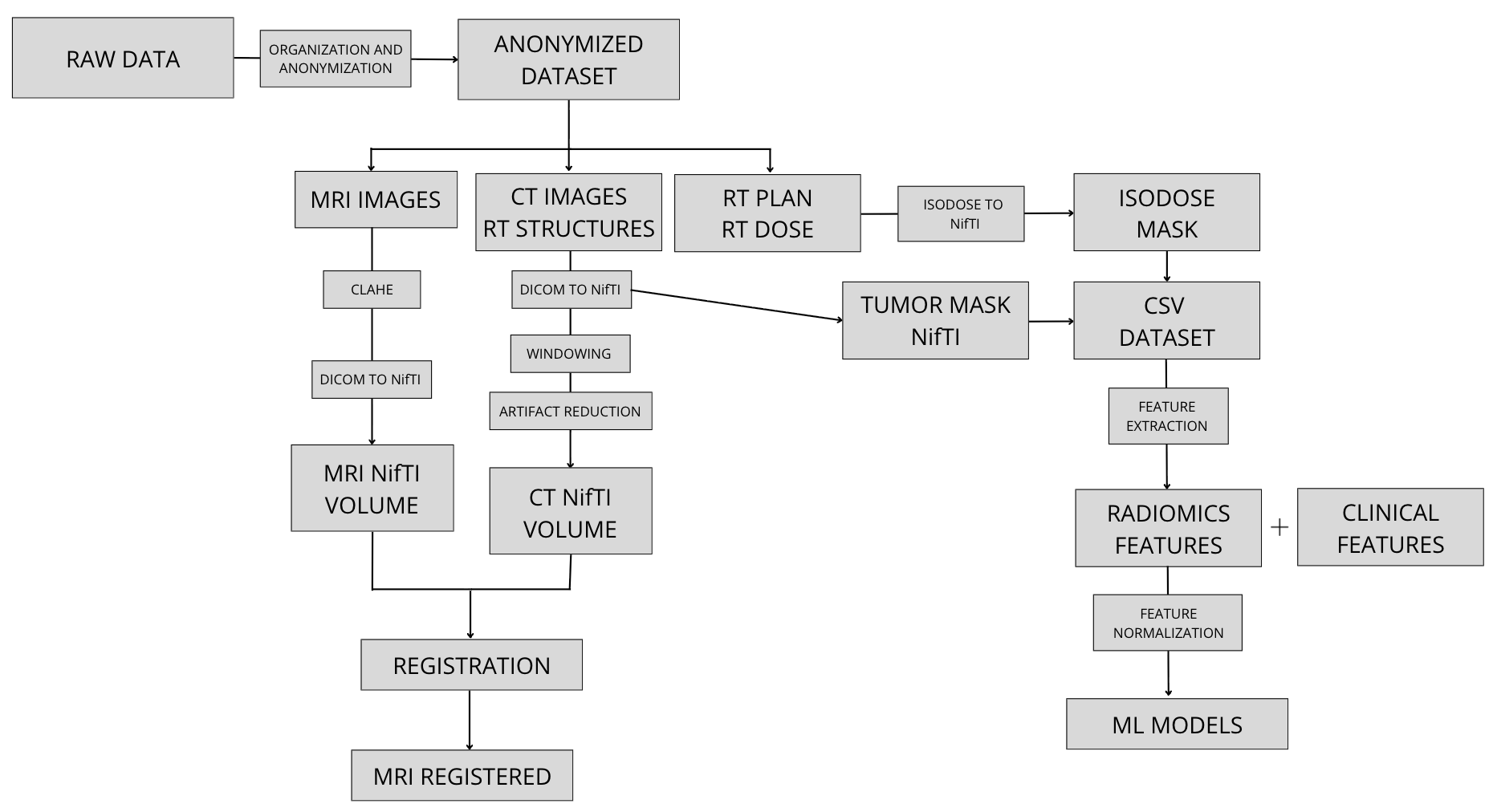}
    \caption{Overview of the Study Workflow.}
    \label{fig:5}
\end{figure}

\subsection{Patient Selection}

The present dissertation examines imaging data from cancer patients with brain metastases undergoing radiotherapy. The dataset under consideration consists of a cohort of 97 patients collected from two institutions, one located in São Paulo, Brazil, and the other in Argentina. The patient population in this study exhibited a range of primary tumor origins and presented with brain metastases. The treatment modalities employed for these patients included LINAC and Gamma Knife techniques. Imaging data were collected during treatment and at subsequent follow-up visits, facilitating the assessment of treatment outcomes and the monitoring of metastatic progression.

\subsection{Data Processing}
\subsubsection{Organization and Anonymization}
\vspace{0.5cm}
The dataset under consideration encompasses both treatment and post-treatment information. The DICOM standard, widely adopted in the field of radiology, serves as the foundational framework for diagnostic imaging. The extension of DICOM into the domain of radiation therapy, designated as DICOM-RT, stands among the inaugural applications of DICOM that transcended the boundaries of diagnostic imaging. This standard facilitates the exchange of medical images and treatment-related information across disparate systems, necessitating the presence of at least one sequence of images alongside corresponding data files.

The RT Structure Set delineates regions of interest (ROIs) pertinent to radiation therapy, including body contours, tumor volumes, and organs at risk. The RT Plan delineates the optimal beam arrangements necessary to attain a desired dose distribution. The RT Dose object is a representation of the actual radiation dose distribution, which is typically visualized through isodose lines expressed as percentages or absolute units (grays). The RT Image, while aligned with aspects of the conventional DICOM framework for radiology, extends beyond image storage to include details on image presentation \cite{ref20}.

To ensure the confidentiality of patient information, the dataset utilized in this dissertation underwent a comprehensive anonymization procedure. The removal of all identifiable tags from MRI and CT images, as well as from dose, plan, and structure files, was a crucial step in the analysis. The names of patients and other sensitive identifiers must be kept confidential.

\subsubsection{Image Registration}
\vspace{0.5cm}
Image processing was performed to prepare the CT and MRI datasets for analysis, ensuring sufficient quality for subsequent procedures. CT images were converted to NIfTI format and subjected to windowing to adjust grayscale values and emphasize anatomical structures \cite{ref11}. Artifact reduction techniques were then applied to improve image clarity. MRI datasets were processed using Contrast Limited Adaptive Histogram Equalization (CLAHE) to enhance local contrast and overall uniformity \cite{ref12}, followed by conversion to NIfTI format. In cases where CT scans were unavailable, the RT structure file from the MRI dataset was converted to NIfTI and processed with CLAHE. All datasets were spatially aligned within a common coordinate framework.

Since the imaging data originated from different sources and coordinate systems, image registration was a crucial step. Registration aligns images to improve anatomical and functional correspondence and is widely applied in multimodal image fusion, image subtraction, treatment planning, computer-aided diagnosis, monitoring, surgical simulations, atlas construction, and radiation therapy \cite{ref13}.

During treatment planning, MRI scans were registered to CT datasets to ensure precise alignment. MRI was primarily used to delineate organs at risk (OARs) and metastatic lesions, while CT served as the reference for radiation dose distribution. This integration enabled optimal dose delivery to tumor volumes while minimizing exposure to surrounding healthy tissues. After segmentation, MRI datasets were transformed into the CT coordinate space for accurate dose computation using the Brainlab Elements\textsuperscript{\textregistered} software.

Due to the superior soft-tissue contrast of MRI, the MRI datasets were defined as the fixed images, while CT datasets served as the moving images. The software computed transformation matrices and applied their inverses to align the MRI and CT images spatially.

Registration accuracy was evaluated using MRI and CT images from the initial and first follow-up scans of 10 randomly selected patients (Patients 54, 59, 61–67). This Subset Selection dataset was used to compare alternative registration tools against those previously employed. Earlier analyses of FSL, 3D Slicer, ANTs and Brainlab Elements identified Brainlab Elements as the most reliable due to its automatic registration and inverse matrix extraction. In this study, additional strategies were tested using SimpleITK and the direct application of inverse transformation matrices.

\subsubsection{Isodose Masks}
\vspace{0.5cm}
The isodose masks were generated by converting RT Dose files into NIfTI format. The volumes of each metastasis were then extracted using DICOMPLYER, an open-source radiation therapy research platform based on the DICOM standard, and exported to a CSV file. The corresponding dosage prescriptions were linked to the respective brain metastases.

Patients were categorized into two groups prior to conversion:

\begin{enumerate}
    \item those with a single dose and plan file, and
    \item those with multiple dose and plan files.
\end{enumerate}

This grouping streamlined file processing and facilitated information extraction from DICOMPLYER. For each patient, the type of treatment was first identified to determine the tumors used in isodose mask creation: for Gamma Knife, Gross Tumor Volumes (GTVs) were selected, whereas for LINAC, Planning Target Volumes (PTVs) were used.

After identifying the relevant tumors, the RT dose was matched to each tumor, and the corresponding volumetric and dosimetric data were consolidated in the CSV file. This file contained the metastasis volume and the mean dose administered. The prescription dose recorded in the RT dose files was compared to the mean dose to determine the effective prescription delivered to each metastasis.

A radius of incidence was then defined, extending 50\% beyond the tumor volume to account for voxel size and margin variability. When the tumor margin was less than one voxel, it was approximated to one voxel to ensure consistent tumor processing.

\subsubsection{Masks Resizing}
\vspace{0.5cm}
To maintain dimensional consistency across all masks—including isodose and tumor masks—they were resized to match the dimensions of the CT images from the initial treatment and MRI images from the first follow-up. However, one patient was excluded due to anomalies encountered during image processing and resizing.

\subsubsection{Patient Clinical Characteristics}
The demographic and clinical characteristics that were considered in this study included the patient's gender, the total number of metastases being treated during the initial treatment, the machine utilized in the initial treatment, the primary tumor location and the decision of treatment. Table~\ref{tab:clinical_data} presents the clinical characteristics of the subjects under consideration. The exclusion of patients 55, 56, 57 and 70 was necessitated by the paucity of data regarding their clinical features. To avoid any potential bias in the algorithm's outcome, it was determined that their removal would be the most judicious course of action, given that a total of 53 patients were to be utilized. The inclusion of features such as the date of the first treatment, the initial follow-up, and, consequently, the calculated interval in months between these two events has been established and independently analyzed, although their formal evaluation within the models has not yet been conducted.
\begin{table}
\caption{Patient demographics and clinical characteristics.}
\centering
\label{tab:clinical_data}
\begin{tabular}{|l|c|l|c|}

\hline
\multicolumn{4}{|c|}{\textbf{Brain Metastasis Patients (N = 53)}} \\ 
\hline
\textbf{Characteristics} & \textbf{Patients (\%)} & \textbf{Characteristics} & \textbf{Patients (\%)} \\
\hline
\multicolumn{4}{|c|}{\textbf{Sex}} \\ \hline
Male   & 15 (28.30\%) & Female & 37 (69.81\%) \\
Others & 1 (1.89\%)   &       &              \\
\hline
\multicolumn{4}{|c|}{\textbf{Number of Metastases}} \\ \hline
$\geq 5$ & 22 (40.74\%) & $< 5$ & 31 (57.41\%) \\
\hline
\multicolumn{4}{|c|}{\textbf{Machine Used}} \\ \hline
LINAC       & 28 (52.83\%) & Gamma Knife & 25 (47.16\%) \\
\hline
\multicolumn{4}{|c|}{\textbf{Primary Tumor}} \\ \hline
Lung                 & 23 (43.40\%) & Breast          & 19 (35.85\%) \\
Kidney               & 2 (3.77\%)   & Colon           & 2 (3.77\%)   \\
Clivus Chordoma      & 1 (1.89\%)   & Pineal Tumor    & 1 (1.89\%)   \\
Intraventricular Tumor & 1 (1.89\%) & Metastatic Melanoma & 1 (1.89\%) \\
Thyroid              & 1 (1.89\%)   & Low Grade Glioma & 1 (1.89\%) \\
Carcinomatous Meningitis & 1 (1.89\%) &  &   \\
\hline
\multicolumn{4}{|c|}{\textbf{Decision after 1st Follow-Up}} \\ \hline
2nd Treatment & 45 (84.91\%) & Follow-Up & 8 (15.09\%) \\
\hline
\end{tabular}
\end{table}

The interval between the first treatment and the initial follow-up (in months) was analyzed as a potential variable for future studies. Follow-up schedules vary according to clinical and individual factors, such as cancer type, treatment modality, and overall patient condition. In clinical practice, follow-ups are usually more frequent in the first years after treatment, when recurrence risk is highest, allowing timely detection of complications and disease progression \cite{ref22}.

\begin{figure}
    \centering
    \includegraphics[width=0.7\linewidth]{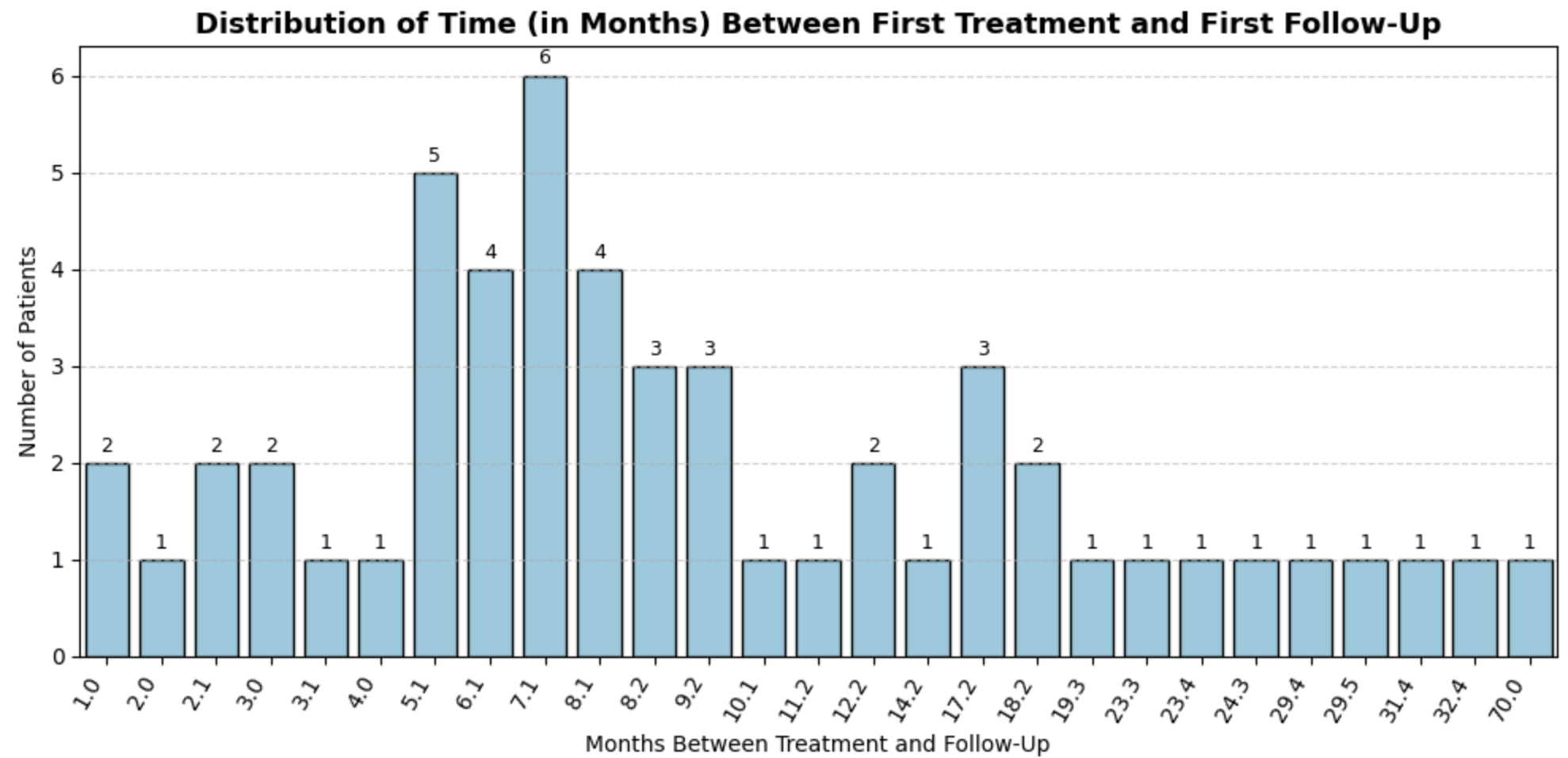}
    \caption{Time interval (months) between first treatment and first follow-up.}
    \label{fig:fu}
\end{figure}

The histogram ~\ref{fig:fu} shows the distribution of follow-up intervals for the analyzed patients. Most follow-ups occurred within 5–8 months after treatment, peaking around 7 months, reflecting the common practice of scheduling reassessments approximately half a year post-treatment.

A smaller subset of patients had early follow-ups (1–3 months), possibly due to clinical complications or higher-risk profiles, while a few exhibited longer intervals (>18 months), including one outlier near 70 months, likely due to irregular adherence or logistical constraints.

Overall, the distribution reveals a central tendency around 6–8 months, consistent with standard brain metastasis monitoring protocols, though the observed dispersion suggests inconsistencies in follow-up timing that could affect long-term analyses \cite{ref23}.

\subsubsection{Radiomics-based ML models}
The final machine learning (ML) model aimed to predict whether a patient undergoing radiotherapy would require additional treatment after the initial follow-up. Five supervised learning algorithms were tested: Random Forest, AdaBoost, XGBoost, SVM, and Decision Tree.

Model development involved patient-specific analysis, computing delta-radiomics features as the absolute difference between features from the isodose and ROI masks. This enabled assessment of metastasis recurrence and the need for re-irradiation. The outcome variable, SI, was binary—0 indicating no further treatment, and 1 indicating the need for re-irradiation.

Only SI was retained as the clinical outcome, while ST was excluded due to limited predictive value. The resulting confusion matrix reflected binary classification (0 vs. 1), emphasizing the importance of minimizing false negatives—patients incorrectly predicted not to need re-treatment—and managing false positives that could lead to unnecessary care. Table ~\ref{tab:confusion_matrix} offers a visual representation of the confusion matrix considering the SI values. 
\begin{table}
\centering
\caption{Confusion matrix for the prediction of subsequent irradiation (SI).}
\begin{tabular}{|c|c|c|}
\hline
\diagbox[width=6em]{\textbf{Predicted}}{\textbf{Actual}} & \textbf{SI = 0 (No 2nd Treatment)} & \textbf{SI = 1 (Requires 2nd Treatment)}  \\ 
\hline
\textbf{SI = 0} & 
\begin{tabular}[c]{@{}c@{}} Patient correctly identified \\ as not requiring re-irradiation\end{tabular} & 
\begin{tabular}[c]{@{}c@{}} Patient not predicted as requiring \\ re-irradiation, but it was necessary\end{tabular} \\ \hline
\textbf{SI = 1} & 
\begin{tabular}[c]{@{}c@{}} Patient predicted to require \\ re-irradiation, but not necessary\end{tabular} & 
\begin{tabular}[c]{@{}c@{}} Patient correctly identified \\ as requiring re-irradiation\end{tabular} \\ 
\hline
\end{tabular}
\label{tab:confusion_matrix}
\end{table}

Before using the final dataset, preliminary model tests were performed, but results were suboptimal due to data imbalance and limited sample size. To address this, new data were added, and relevant clinical features, such as gender, age, primary tumor site, treatment machine, and number of metastases—were incorporated, as these variables have shown strong correlations with local control in previous studies \cite{ref24} \cite{ref25} \cite{ref26}.

The cohort included 53 patients with 415 metastases, though some cases were excluded after normalization inconsistencies were detected. The dataset was split 80/20 into training (141 cases, 1,339 features) and testing (36 cases).

Categorical features were encoded using one-hot encoding to prevent artificial ordinal bias and maintain interpretability. Five supervised algorithms, XGBoost, Random Forest, AdaBoost, Decision Tree, and SVM, were trained and optimized through RandomizedSearchCV with 5-fold cross-validation. Each model was evaluated for its predictive performance in identifying recurrence of brain metastases.

\section{Results}

\subsubsection{Model Performance}

Using the optimal parameters identified in the previous section, new results were generated with the test set. Table~\ref{tab:performance} summarizes and compares the performance of the five machine learning models, providing a clear overview of their predictive capacity for forecasting brain metastasis recurrence.

\begin{table}
\centering
\caption{Performance metrics of the models on the training and test sets.}
\begin{tabular}{|l|c| c|c |c|c| c|c |c|}
\hline
\multirow{2}{*}{\textbf{Model}} & \multicolumn{2}{c|}{\textbf{Precision}} & \multicolumn{2}{c|}{\textbf{Recall}} & \multicolumn{2}{c|}{\textbf{F1-Score}} & \multicolumn{2}{c|}{\textbf{Accuracy}} \\ \cline{2-9}
 & \textbf{Train} & \textbf{Test} & \textbf{Train} & \textbf{Test} & \textbf{Train} & \textbf{Test} & \textbf{Train} & \textbf{Test} \\ \hline
XGBoost        & 1.00 & 0.93 & 1.00 & 0.93 & 1.00 & 0.93 & 1.00 & 0.94 \\ \hline
Random Forest  & 1.00 & \textbf{0.98} & 1.00 & \textbf{0.94} & 1.00 & \textbf{0.96} & 1.00 & \textbf{ 0.97} \\ \hline
AdaBoost       & 1.00 & \textbf{0.98} & 1.00 & \textbf{0.94} & 1.00 & \textbf{0.96} & 1.00 & \textbf{0.97} \\ \hline
SVM            & 1.00 & 0.85 & 1.00 & 0.76 & 1.00 & 0.79 & 1.00 & 0.86 \\ \hline
Decision Tree  &  1.00 & 0.88 & 0.99 & 0.91 & 0.99 & 0.89 & 0.99 & 0.92 \\ \hline
\end{tabular}
\label{tab:performance}
\end{table}

As shown in the Table 3, all models achieved excellent training performance (precision, recall, F1-score, and accuracy near 1.00), suggesting strong learning but also potential overfitting.

On the independent test set, the models’ differences became more evident. Random Forest and AdaBoost achieved the best overall results, with accuracies of 0.97, followed by XGBoost (0.94), Decision Tree (0.92), and SVM (0.86). The superior performance of Random Forest aligns with findings from \textit{Munai et al.}, \textit{Yu et al.}, and \textit{Li et al.}, who also reported its robust predictive capabilities in similar clinical contexts.

In terms of precision, Random Forest and AdaBoost (0.98) outperformed XGBoost (0.93), while Decision Tree (0.88) and SVM (0.85) showed higher false-positive rates. For recall, Random Forest, AdaBoost, and XGBoost (>=0.93) demonstrated strong sensitivity, whereas SVM (0.76) performed weakest. Considering the F1-score, which balances both metrics, Random Forest and AdaBoost again led (0.96), followed by XGBoost (0.93), Decision Tree (0.89), and SVM (0.79).

In summary, while all models performed well on training data, Random Forest and AdaBoost provided the best generalization, achieving the most balanced performance across all evaluation metrics. The detailed confusion matrices for these two models are presented below.

\begin{figure}
    \centering
    \includegraphics[width=0.7\linewidth]{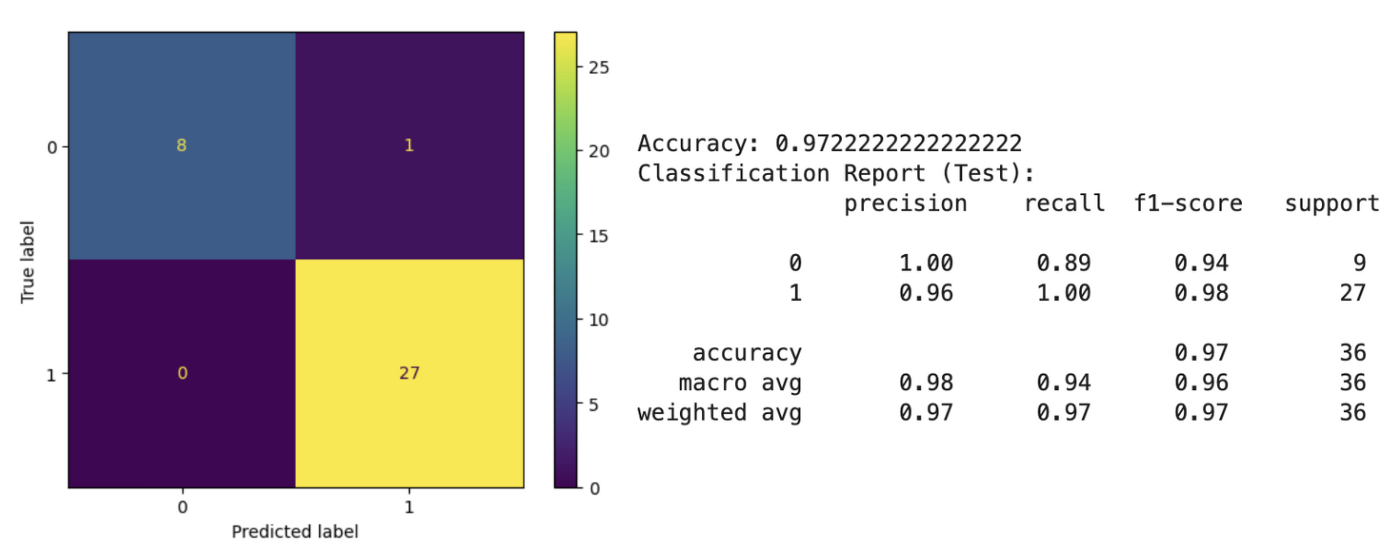}
    \caption{Confusion matrix and performance metrics for the Random forest model on the test set.}
    \label{fig:1}
\end{figure}

\begin{figure}
    \centering
    \includegraphics[width=0.7\linewidth]{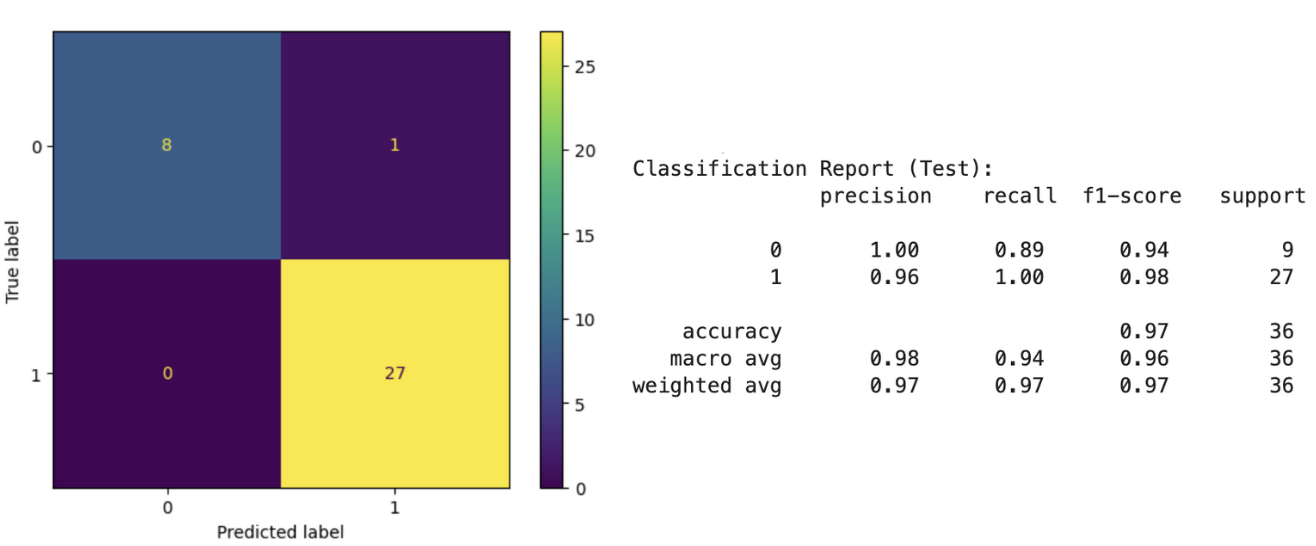}
    \caption{Confusion matrix and performance metrics for the Adaboost model on the test set.}
    \label{fig:2}
\end{figure}

For the models that achieved the best results, the most relevant features were identified using the feature\_importances\_ method with the weight-based importance criterion. This approach evaluates the contribution of each feature by averaging its impact across both labels. From this analysis, the eight most relevant features were selected. Both of the best-performing models identified the same feature as the most important. This convergence reinforces the robustness of the findings, as independent algorithms arriving at the same conclusion provide stronger evidence for the relevance of that feature.

\begin{figure}
    \centering
    \includegraphics[width=0.7\linewidth]{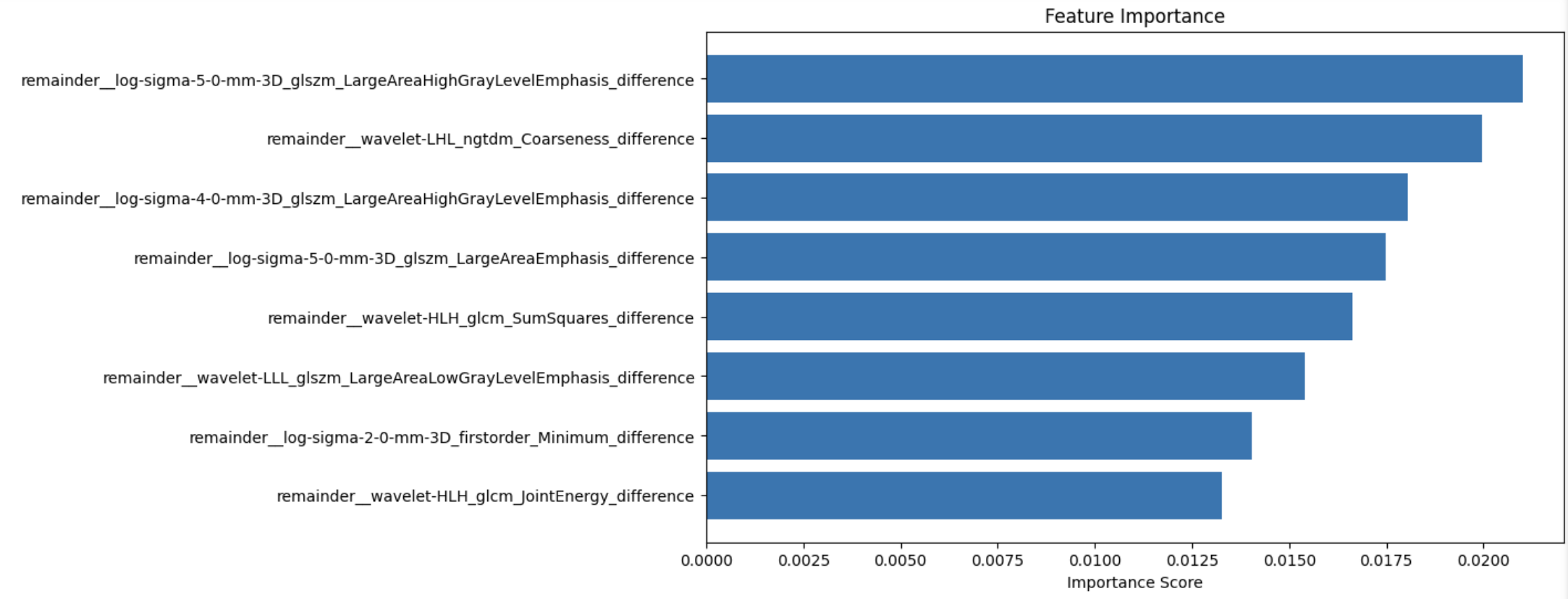}
    \caption{Top 8 Most Important Features for Random forest model. }
    \label{fig:3}
\end{figure}

\begin{figure}
    \centering
    \includegraphics[width=0.7\linewidth]{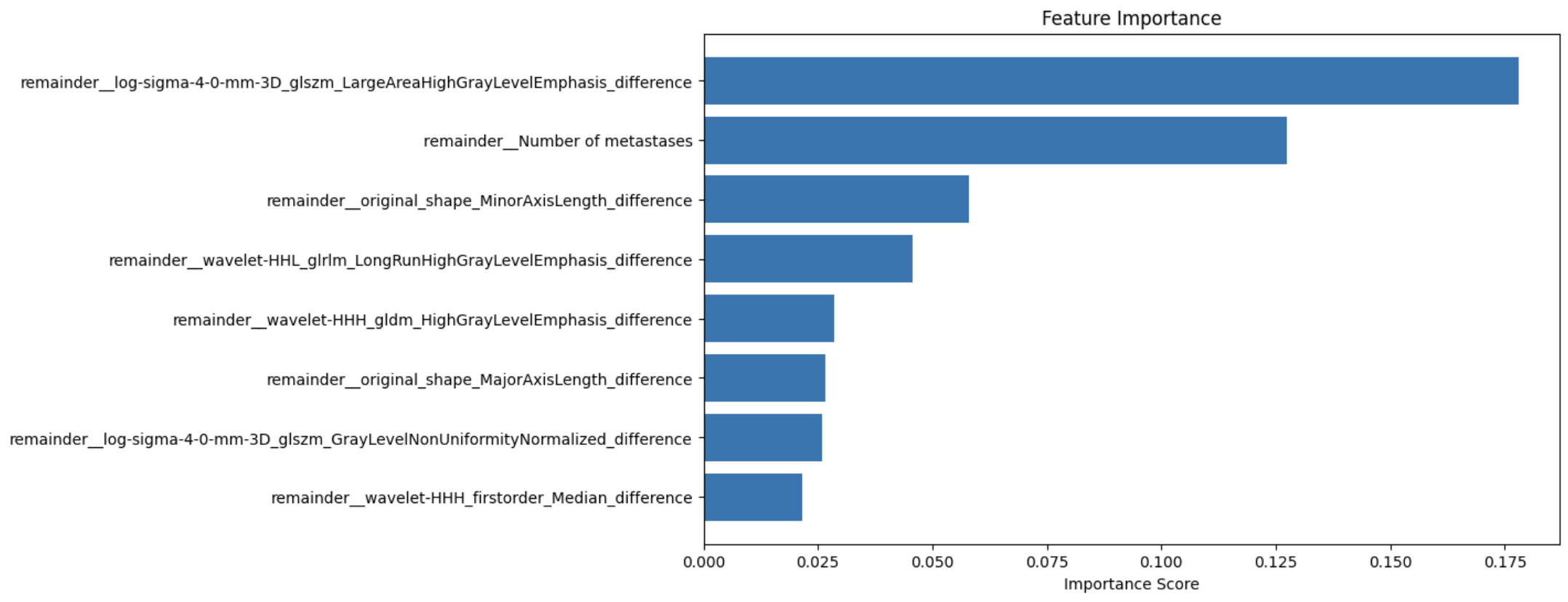}
    \caption{Top 8 Most Important Features for Adaboost model.}
    \label{fig:p4}
\end{figure}

The results show that the top models identified the same key feature. This supports the findings, as different algorithms pointing to the same feature provide stronger evidence.

\section{Discussion \& Conclusion }
This study demonstrates that the integration of radiomic and clinical features plays a central role in the development of machine learning models aimed at predicting brain metastasis recurrence. The radiomic pipeline, encompassing preprocessing, registration, segmentation, and feature extraction, revealed how strongly these features depend on image quality, alignment, and tumor anatomy. Registration inconsistencies, resampling errors, and the presence of extremely small tumors resulted in unstable or missing radiomic features, ultimately reducing the usable dataset and affecting model robustness. This outcome aligns with existing literature indicating that radiomics is highly sensitive to acquisition parameters, segmentation precision, and inter-modality harmonization.

Clinical features, in contrast, demonstrated greater stability and consistency, providing meaningful context to complement radiomic descriptors. Variables such as primary tumor site, number of metastases, patient sex, and treatment modality contributed relevant biological and treatment-related information. However, incomplete clinical records led to additional patient exclusions, reinforcing that clinical data, while more structured, must be rigorously documented to maximize predictive utility.

The model comparison highlighted that ensemble methods, particularly Random Forest and AdaBoost, were best suited for integrating heterogeneous radiomic and clinical inputs. Their strong training and test set performance suggests that combining both feature domains enhances the model’s capacity to identify recurrence-related patterns. Nonetheless, the limited dataset size and significant class imbalance restricted generalization potential, emphasizing the need for larger cohorts, improved preprocessing pipelines, and harmonization strategies to strengthen model reliability.

Overall, the combined analysis of radiomic and clinical features underscores their complementary nature: radiomics captures quantitative imaging traits, while clinical variables provide biological and contextual depth. Despite the challenges encountered, this study supports the feasibility of multi-feature predictive modeling for brain metastasis recurrence and highlights clear avenues for future research. By expanding datasets, improving imaging pipelines, and refining feature stability, future models may achieve higher accuracy and clinical applicability, contributing to more personalized and effective management of patients undergoing radiotherapy for brain metastases.

%

%
%
%
%

\end{document}